\documentclass[12pt]{article}
\pdfoutput=1
\usepackage[italian,english]{babel}
\usepackage[utf8]{inputenc}
\usepackage[T1]{fontenc}
\usepackage{hyperref} 
\usepackage{amsmath,amssymb}
\usepackage{graphicx}
\usepackage{sidecap}
\usepackage[affil-it]{authblk}

\begin{document}
\title{Applications of Biological Cell Models in Robotics}
\author{
	Michele Braccini\\
	m.braccini@unibo.it
}

\affil{
	Department of Computer Science and Engineering (DISI)\\
	Campus of Cesena\\
	{\em Alma Mater Studiorum} Universit\`a di Bologna,
	Cesena, Italy
}
	
\date{}

\maketitle

\begin{abstract}
In this paper I present some of the most representative biological models applied to robotics.
In particular, this work represents a survey of some models inspired, or making use of concepts, by gene regulatory networks (GRNs): these networks describe the complex interactions that affect gene expression and, consequently, cell behaviour.
\end{abstract}

\section{Introduction}\label{Introduction}
In order to evolve robots, or sets of robots, capable of increasingly complex tasks, we need to apply to the robotics field more and more powerful models, techniques and methodologies. 
Natural systems exhibit properties like robustness, adaptiveness, flexibility, scalability and reliability; and they represent a source of interest for the construction of artificial systems.
In particular, natural systems like insect colonies and flocking birds exhibit intelligent emergent collective behaviours.
We are interested in the dynamical mechanisms at the basis of these systems that lead to the creation of such global level structures, like self-organisation behaviours, from interactions among lower-level components.
Complex system science deals with the study of how these low-level parts of a system give rise to the collective behaviours and how the system interact with its environment.
Some examples of complex systems are the ant colonies, the human brain, the cell, the society and the ecosystem.
Real living cells, especially, are both robust and adaptive (can maintain their functions in spite of noise and they are able to adapt to new environmental conditions) and they are considered critical systems: theirs dynamics is between the order and chaos.
It has been hypothesized that the complexity lies at the edge of order and chaos and so the biological concepts and mechanisms underlying of the living cells are exploitable, hopefully, to design agent's behaviours as complex as these.
In particular the biological genetic regulatory networks (GRN in short) model the interaction and dynamics among genes and they are considered complex dynamical systems able to produce a wide diversity of living cells and organisms.
These GRN can be engineered and used to control or to evolve robots.
In this paper I survey some of the most relevant examples of adoption of GRN models in robotics.
In certain case the dynamics of the GRN-based models are used to directly control robots, in others the GRN mechanisms are adopted, similarly to the biological morphogenesis, to develop the robot's neural network control, the robot's morphology or the pattern formation for swarm of robots.

The paper is organized as follows.
In section 2 I present a brief introduction, at high level of abstraction, of the gene regulation, useful to introduce the concepts that will be used by the models presented later.
Some models based on genetic regulatory networks, found in the literature, are introduced in section 3, grouped to reflect how these models are employed in robotics.
In section 4 it is given an introduction, taken from \cite{jin2011morphogenetic}, to the Morphogenetic Robotics.
The last section provides a conclusion to this work.

\section{Regulation of Gene Expression}\label{grn}

\paragraph{DNA}
DNA (\textit{deoxyribonucleic acid}) is a nucleic acid that carries most of the genetic instructions.
The DNA is what of which genes are compound.
Genes are the hereditary units that transmit information from parents to offspring (they contain the information necessary for the production of a protein or RNA).
Within cells, DNA is organized into long structures called chromosomes.
Each chromosome is composed of a very long DNA molecule along which are arranged hundreds or even thousands of genes.
When a cell is preparing to divide, the DNA of its chromosomes is duplicated so that each daughter cell gets an identical set of genes.
In each cell, the genes arranged along the DNA molecules encoding the information to build other molecules of the cell. In this way the DNA controls the development and maintenance of the whole organism \cite{biologia}.

\begin{figure}[h]
	\centering
	\includegraphics[scale=0.8]{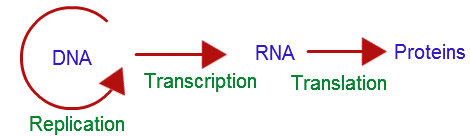}
	\caption{Central Dogma of Molecular Biology}\label{fig:1}
\end{figure}

\paragraph{Central dogma of molecular biology}
The \textit{central dogma of molecular biology} is an explanation of the flow of genetic information, from DNA to RNA, to make a functional product (a protein), within a biological system and it was first stated by Francis Crick in 1956.
A simple representation of the central dogma is observable in Figure \ref{fig:1}.

The central dogma suggests that DNA contains the information necessary to make all of our proteins, and that RNA is a messenger that carries this information to the ribosomes.
Ribosomes are complex molecular machines and they are within all living cells; they serve as the site of \textit{biological protein synthesis} (translation) \footnote{In translation messenger the RNA (mRNA), produced by transcription from DNA, is decoded by a ribosome to produce a specific amino acid chain, or polypeptide. The polypeptide later folds into an active protein and performs its functions in the cell.}. 
The process by which the DNA instructions are converted into the functional product is called \textbf{gene expression}, and has two key stages: \textit{transcription} and \textit{translation}.
The central dogma states that the pattern of information that occurs most frequently in our cells is:
\begin{itemize}
	\item From existing DNA to make new DNA (DNA replication, that is the process by which a cell makes an identical copy of its genome before it divides.)
	\item From DNA to make new RNA (transcription)
	\item From RNA to make new proteins (translation).
\end{itemize}
With modern research it is becoming clear that some aspects of the central dogma are not entirely accurate.
Current research is focusing on investigating the function of non-coding RNA (a functional RNA molecule that is transcribed from DNA but is not translated into a protein) \footnote{\url{http://www.yourgenome.org/facts/what-is-the-central-dogma} Date:04/01/2016}.
Non-coding RNAs are involved in many cellular processes, one of these is the \textbf{gene regulation}.

\paragraph{Gene Regulatory Network}\label{grn}
All steps of gene expression can be modulated, since passage of the transcription of DNA to RNA, to the post-translational modification of the protein produced. 
Hence, gene expression is a complex process regulated at several stages in the synthesis of proteins.
In addition to the DNA transcription regulation, the expression of a gene may be controlled during RNA processing and transport (in eukaryotes), RNA translation, and the post-translational modification of proteins.
This gives rise to \textit{genetic regulatory systems} structured by \textit{networks of regulatory interactions} between DNA, RNA, proteins and other molecules \cite{de2002modeling}: a complex network termed as a \textbf{gene regulatory network (GRN)}.
Some, noteworthy, kind of proteins are the \textbf{transcription factors} that bind to specific DNA sequences in order to regulate the expression of a given gene. 
The power of transcription factors resides in their ability to activate and/or repress transcription of genes. 
The activation of a gene is also referred to \textit{positive regulation}, while the \textit{negative regulation} identifies the inhibition of the gene.

The \textbf{regulation of gene expression} is essential for the cell, because it allows it to control their internal and external functions.
Furthermore, in multicellular organisms, gene regulation drives the processes of \textbf{cellular differentiation} and \textbf{morphogenesis}, leading to the creation of different cell types that possess different gene expression profiles, and hence produce different proteins/have different ultrastructures that suit them to their functions (though they all possess the genotype, which follows the same genome sequence) \footnote{\url{https://en.wikipedia.org/wiki/Regulation_of_gene_expression} Date:03/01/2016}.
Therefore, with few exceptions, all cells in an organism contain the same genetic material \cite{de2002modeling}, and hence the same genome (the haploid set of chromosomes of a cell). 
The difference between the cells are emergent and due to regulatory mechanisms which can turn on or off genes.
Two cells are different, if they have different subsets of active genes \cite{eggenberger1997evolving}.

\section{GRN-based Models and Applications in Literature}\label{grn application}
In this section I present some examples of models based on the mechanisms of the genetic regulatory networks applied to the field of robotics.
Below, the examples are grouped so as to reflect how the models are used in robotics: automatically evolve a robot controller, automatically design a robot morphology, generate pattern for swarm robotics and automatically co-evolve morphology and controller for robot.

\subsection{GRN-based Models for Designing Robot Control}
	
\subsubsection{First Example} \label{cell_eggenberger}
Eggenberger in the paper ``Cell Interactions as a Control Tool of Developmental Processes'' \cite{eggenberger1996cell} suggests that biological concepts as developmental processes are useful and applicable to the field of evolutionary robotics.
With the proposed model the length of the genome can be reduced, because no explicit data about the connectivity pattern of the neural net are stored in the genome.
The connectivity is not directly encoded in the genome itself but it's mainly determined by the developmental processes.
The artificial evolutionary system (AES) includes the following biological concepts and mechanisms:
\begin{itemize}
	\item Regulatory Units and Transcription Factors, Cell Adhesion Molecules (CAM) and Cell Receptors;
	\item Cell Differentiation;
	\item Cell Division;
	\item Cell Adhesion.
\end{itemize}
The artificial genome is implemented as a string of integers and it is composed by \textit{regulatory units} and \textit{structural genes}.
Regulatory Units are used to activate or inhibit the activity of the structural genes, the latter (if active) modulate the developmental processes producing a substance among these four: transcription factors, cell adhesion molecules, receptors or artificial functions (class that is used to define whether a cell should divide or not).
\paragraph{Regulation of Gene Activity}  If a cell contains a transcription factor, its code is compared with the code of all regulatory unit in this cell. Depending on a defined \textit{affinity function} the regulatory unit is the activated or inhibited. If a regulatory unit is activated also its structural genes will be activated.
\paragraph{Cell Differentiation} Two cells are different if they contain different subset of active genes in the genome. The implemented mechanisms to obtain different cells are: \textit{cell lineage} and \textit{cell induction}.
The cell lineage is an autonomous mechanism in which cell differentiation depends on intracellular factors, which are unevenly distributed in different cells.
In the cell induction the cells become different because they get different signaling from other cells.
To simulate this mechanism the author implemented three different pathway to exchange information between cells: first, there are substances which don't leave the cell and which regulate the activity of genes; second, there are substances which can penetrate the cell wall and activate all cells which are near by; third, there are specific \textit{receptors} on the cell surface which can be stimulated by substances.
If a transcription factor has a high enough affinity to the receptor, a gene or group of genes is turned on or off.
Only those cells which have a specific receptor on the cell surface will respond to a certain substance.
After the process of cell differentiation is finished, the different active genes will determine which substances are produced in a cell.
\paragraph{Cell Division} The proposed model is able to simulate cell growth. If the structural genes for cell division in a cell is active, the cell divides itself.
The gene activity is dependent, in addition to the affinity function, also on the concentration of the transcription factor.
At a certain moment, due to its increased concentration, the transcription factor will turn off the gene for cell division and the growth will stop. 

This model has been used to evolve a neural control structure for an autonomous agent.
The artificial neurons are the standard ones, with a sigmoidal activation function.
As cells can become different, they will express also different substances. 
To connect two cells or neurons, there are two different types of adhesion molecules and these are stored in lists in the cell.
The members of the first list of one neuron are compared with the list of another neuron: if two adhesion molecules of the two different lists have a high affinity to each other a link from the first cell to the second cell is established (if two or more links to the same cell are possible, the substance with the greatest affinity is chosen).
The developed neural network has to be linked to the sensors and motors of a real robot, and in order to leave this task to the algorithm implemented, Eggenberger has defined sensory and motors cells with a list of adhesion molecules, in this way other cells can connect to them.

In this paper two experiments are presented and the neural controller is evolved by means of a genetic algorithm that is at the basis of the AES.
In the first experiment the robot has to accomplish an object avoidance task (the corresponding fitness is increased if the robot sees an object but avoids a collision) and the second task is a phototaxis plus object avoidance (in this case the fitness is the same of the first task, in addition is increased if the robot moves away from its initial position and if the robot is near the light source).
For both tasks the number of initial cells are the same (thus the length of the genome is fixed) and by means of the mechanisms introduced in the AES the neurons can grow and multiply.
Therefore, the neural network is evolved and the number of cells is multiplied even though the genome length was fixed.

The purpose of this work is to show that the introduced AES, with the gene regulatory mechanisms, can control the main developmental processes and can evolve functioning neural networks for autonomous agents with number of neurons and patterns of connections not explicitly stored in the artificial genome.

\subsubsection{Second Example}\label{morphogenesis_nn}
Another example of design of neural network controller, in addition to that previously presented, can be found in the following papers: ``Morphogenesis of neural networks'' \cite{michel1995morphogenesis} and ``From the Chromosome to the Neural Network'' \cite{michel1995chromosome}, both of the same authors.
In these papers it is proposed a model, inspired from biology, of morphogenesis process with the aim to synthesize an artificial neural network to lead an autonomous robot.
Both structure and weights of the neural network are defined 	by the morphogenesis process.
The model was inspired to the biological principle of the \textit{proteins synthesis regulation}
\footnote{Protein synthesis is the final stage of gene expression. 
	Once synthesized, most proteins can be regulated in response to extracellular signals and in addition, the levels of proteins within cells can be controlled by differential rates of protein degradation. \url{http://www.ncbi.nlm.nih.gov/books/NBK9914/}}.
The morphogenesis process starts on a single cell enclosing a chromosome and a message list; this list corresponds to the set of proteins available in the biological cell.
Cells can divide and establish connections among them by means of a sort of production system that uses and produces messages (representing proteins), through rules.
Each rule can be divided in three parts, two conditions and one action part: a set of messages whose absence in the cell message list is necessary for the rule to be fired (corresponding to the biological \textit{repressors}), a set of messages whose presence is necessary for the rule to be fired (corresponding to the biological \textit{activators}) and a set of produced messages which are added to the cell message list when the rule is fired (corresponding to the \textit{biological synthesized proteins}).
This morphogenesis process is general: it is able to create any kind of neural network. 
It allows recurrent connections, different kinds of neurons with different transfer functions and different kinds of links with different learning rules. 
Thus the space of neural networks explored is theoretically unlimited \cite{michel1995chromosome}.

To evolve chromosomes classical genetic algorithms have been used with a single point crossover, random mutations and a genetic operator which can add or remove a random element in one of three parts of a rule, or add or remove whole rule at the end of the chromosome.
Natural selection uses a binary fitness function (life or death); the evolution is progressive and in order to obtain more complex and powerful systems it is necessary evolve the environment (once a large enough population is able to survive, the environment get a bit more hostile, so that only the elite of this population can survive, and so on, until the individuals develop elaborate behaviours).
A simple goal \textit{go towards food} was chosen; a kind of robot \textit{metabolism} was introduced (a variable represents the internal energy of the robot and it is increased each time the robot eats food and decreased when it moves or it is motionless; if it reaches zero the robot ``dies'' and is eliminated from the population).
After around 300 generations of the genetic algorithm, the population of the neural networks evolved were able to produce the desired behaviour, the robots were attracted by food.

This proposed method, inspired mainly to the biological morphogenesis, has demonstrated to be able to produce neural networks (structure and weights) that generate remarkable, even if simple, task (as the attraction by food).

\subsubsection{Third Example}\label{bn_robotics}

\paragraph{Boolean Networks}
Boolean networks (BNs) are a prominent example of complex dynamical systems and they have been introduced by Kauffman \cite{kauffman1969metabolic} as a GRN model: as they are able to reproduce very important phenomena in gene regulation.
From an engineering perspective these models are very fascinating because they can exhibit rich and complex behaviours in spite of the compactness of their description.

A BN is a discrete-state and discrete-time dynamical system whose structure is defined by a directed graph of N nodes, each associated to a Boolean variable $x_i$, $i= 1,\dots,N$, and a Boolean function $f_i(x_{i_{1}}, \dots, x_{i_{K_{i}}})$, where $K_i$ is the number of inputs of node $i$.
The arguments of the Boolean function $f_i$ are the values of the nodes whose outgoing arcs are connected to node $i$.
The state of the system at time $t, t\in\mathbb{N}$, is defined by the array of the $N$ Boolean variable values at time $t$: $s(t)\equiv(x_1(t), \dots, x_N(t))$.
Therefore, the BN models are oriented graphs in which there are the following simplifications: 
\begin{itemize}
	\item each node of the network corresponds to a gene;
	\item the genes are binary devices (ON or OFF);
	\item there is an arc from a node (e.g. A) to an another node (B) if this last gene B is influenced by the activation A.
\end{itemize}
The most studied BN models are characterised by \textit{synchronous} dynamics (nodes update their states at the same instant) and \textit{deterministic} functions; in these, given an initial condition the dynamics of the networks can be described by means of a \textit{trajectory}: sequence of states in consecutive time instants.
However, many variants exist, including asynchronous and probabilistic update rules.\cite{roli2011design}

In the paper ``On the Design of Boolean Networks Robots'' \cite{roli2011design} it was presented the use of Boolean networks for controlling robot's behaviour.
The approach proposed consists in using one or more BNs as robot program so that the robot dynamics can be described in terms of trajectories in a state space.
The authors propose a design methodology based on metaheuristics in which the design of a BN is modelled as a constrained combinatorial optimisation problem: 
the algorithm manipulates the decision variables which encode structure and Boolean functions of a BN. 
A complete assignment to those variables defines an instance of a BN.
This technique uses an evaluator that produces an objective function value that represents the performance of the current BN and the feedback to the metaheuristics algorithm, that, in turn, proceeds with the search.
Another possible way, suggested in this article and that can be combined with the previous presented, to design the BN for a robot program is to exploit its dynamics in order to satisfy given requirements.
For example, the attractors with largest basins of attraction may correspond to the high-level robot's behaviours and the transitions between attractors to the transitions between behaviours.  

The case study presented consists of a robot that must be able to perform two different behaviours: going towards the light (phototaxis) and subsequently moving away from it (antiphototaxis) after perceiving a sharp sound (like an hand clapping).
The environment, in which the robot is simulated and later tested in reality, consists of a square ($1m \times 1m$) with a light source positioned in one corner.
The robot, in the beginning of the experiment, is located in random position close to the opposite corner of the arena with respect to the light and the performance measure used to evaluate the robot behaviour is an error function that has to be minimised (smaller is the error, better is the robot performance).
The BN implementing the robot program is subject to a synchronous and deterministic update and the number of network nodes is has been set to 20 (sensors and actuators have been mapped onto some node of the BN).
The boolean network was designed with a local search techniques, that is a simple stochastic descent in which a move can change one value in a node function's truth table; a random entry in the truth table of a randomly chosen node is chosen and accepted if the corresponding BN has an evaluation not worse than the current one.
The initial connections among nodes are randomly generated with $K=3$ (no self-connections) and are kept fixed during the search; the initial Boolean functions are generated by setting the 0/1 values in the truth tables uniformly at random.
The BN-robot is trained in two sequential phases: in the first, the learning feedback is an evaluation of the robot's performance in achieving only phototaxis and in the second the performance measure takes into account both the phototaxis and antiphototaxis. 
In this way, it become possible to study the properties of the evolution of the BN-robot when its behaviour must be adapted to a new operational requirement.

The results obtained from this experiment were presented in the paper \cite{roli2012preliminary}.
Analysing the dynamics of BN-robots trained, using concepts of dynamical systems theory and complexity science, they have found that the successful performing robots, which show the capability of robustly attaining the learned behaviours while adapting to new tasks to perform, are characterised by both number of fixed points and complexity higher than those of unsuccessful ones.
The number of fixed points is an indicator of the generalisation capabilities of the system as they represent micro-behaviour which are combined to achieve a global behaviour and the measure of the complexity used is the LMC complexity \footnote{LMC complexity is defined as $C = HD$, where $H$ is the \textit{entropy} and $D$ is the \textit{disequilibrium} of the BN states in the trajectories.}.
These results are in accordance with the conjecture that artificial systems able to balance robustness and evolvability work at the border between order and chaos as the living systems, an example are cells.

In the papers ``A Developmental Model for the Evolution of Complete Autonomous Agents'' \cite{dellaert1996developmental} and ``Co-evolving Body and Brain in Autonomous Agents using a Developmental Model'' \cite{dellaert1994co} it is presented a model for neural development in which a random Boolean Network is used as an abstraction of the genetic regulatory network inside a cell. 
The introduced developmental process has showed to be able to successfully evolve agents than can execute simple tasks (i.e. line following).
	
\subsubsection{Others Relevant Examples in the Literature}
Another interesting example of evolution of a neural network controller using biological principles can be found in the article ``Evolving the morphology of a neural network for controlling a foveating retina - and its test on a real robot'' \cite{hotz2003evolving}.
The proposed model combines artificial evolutionary techniques with bio-inspired developmental processes in order to evolve a neural network that acts as an artificial foveating retina (that is, move the ``eye'' in such a way that an incoming peripheral sensory stimulus falls in the center of the eye, the eye has to learn to foveate on the stimulus).
In particular, this system exploits mechanisms like gene regulation and developmental mechanisms like cell division, axonal outgrowth, synaptogenesis and learning for controlling the structure of the neural network and the synaptic weights.
After the simulation, the evolved controller was tested in a real robot arm equipped with a CCD camera and the arm has proved to be able to foveate considerably well.

In the paper ``Harnessing Morphogenesis'' \cite{jakobi2003harnessing} is presented another fascinating example of controller design making use of a biologically inspired developmental model.
This system is able to develop a multicellular organism, starting from a single cell, exploiting similarities with biological morphogenesis.
The behaviour of the cell, during development, is controlled by a GRN that can be thought as a dynamical system.
The product of this developmental process is interpreted as a recurrent neural network robot controller; this model was able to evolve controllers for accomplish a corridor following and an object avoidance task.

The paper ``Evolving Embodied Genetic Regulatory Network-Driven Control Systems'' \cite{quick2003evolving} presents experiments in which a GRN-based controller is embodied in artificial organisms.
In this model, called \textit{Biosys}, the interplay between the dynamics of the embodied GRN controller (the suitable genome is evolved through a genetic algorithm) and the environment gives rise to coherent observable emergent behaviours.
It's presented a successful experiment of a simulated robot, guided by its GRN controller, able to fulfil phototaxis, but the key point of this work it's to remark the importance of the role of the environment in the generation of observed behaviour;
the environment ``selects'' the cell dynamics able to produce the desired behaviour.
	
\subsection{GRN-based Models for Evolving Robot Morphology}\label{evolving_eggenberger}
The paper ``Evolving Morphologies of Simulated 3d Organisms Based on Differential Gene Expression'' \cite{eggenberger1997evolving} reports a biologically inspired model used to evolve 3d shapes of simulated, multicellular organisms.
The model has the same concepts and biological mechanisms introduced in the previously presented article of the same author \cite{eggenberger1996cell}, in addition introduces the \textit{positional information} and pattern formation in development.
With this last mechanism the cells acquire positional identities as in a coordinate system and then interpret this information according to their genetic constitution and developmental history.
An example of such mechanism is a concentration of gradient of a \textit{morphogen} which every cell is able to read.
A morphogen is a substance governing the pattern of tissue development in the process of morphogenesis, and the positions of the various specialized cell types within a tissue \footnote{\url{https://en.wikipedia.org/wiki/Morphogen} Date: 04/01/2016}.
In the implementation proposed by Eggenberger the morphogen is just a kind of transcription factor (TF) which can diffuse to other cells and can change the state of some genes in cells able to read this message.
This mechanism, already implemented by the regulatory mechanism in the AES, is not just a simple signaling, because the reading mechanism (the \textit{cis-regulators}, which are binding sites for transcription factors) is controlled by the AES. 
Therefore the same morphogen can have very different effects on different cell.
Some examples of such effects are changes in cell type, cell division rate or motility.

\begin{SCfigure}[][h!]
	\centering
	\includegraphics[width=0.4\textwidth]{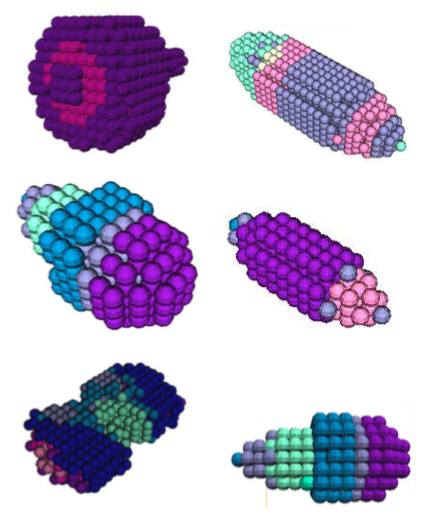}
	\caption{Examples of evolved forms by means of the AES; the fitness function evaluated only the number of the cells and the bilaterality of the found organisms. Image taken from \cite{eggenberger1997evolving}.}\label{fig:egg_3d}
\end{SCfigure}
Using the proposed AES, Eggenberger was able to evolve three dimensional shapes that could be used as projects for three-dimensional robot.
Some examples, taken by the original paper \cite{eggenberger1997evolving}, of these evolved forms are represented in Figure \ref{fig:egg_3d}.

\subsection{GRN-based Models for Pattern Generation in Swarm Robotics}\label{network_motifs}
Inspired by the biological morphogenesis and the evolution and structure of \textbf{networks motifs}, in the paper ``Evolving Network Motifs based Morphogenetic Approach to Self-Organizing Robotic Swarms'' \cite{meng2012evolving} it's presented a GRN-based control model.
This model has the aim to autonomously generate dynamic pattern for swarm robot in complex environment.
Network motifs are pattern of interconnections occurring in complex networks at numbers that are significantly higher than those in randomized networks \cite{milo2002network}; therefore they represent building blocks for most complex networks.
The authors propose a developmental method where the artificial GRN-based controller will be automatically evolved by an evolutionary algorithm using some predefined network motifs as basic building blocks.

The aim of this model is to generate suitable shapes so that swarm of robots (with limited sensing and communication capabilities) can traverse an unknown environment with various constraints. 
Inspired by the biological morphogenesis, in which the morphogen gradients are either obtained from the mother cells or generated by a few cells known as organizers, an organizing robot is selected in order to generate the final target shape (considering the current environmental constraints) for the swarm robot.
The regulation of gene expression is used to model the base concepts of the general GRN-framework.
This framework will be embedded into each robot of the system, but only the organizing robots will activate the framework and generate the suitable shapes virtually in its own mind.
Then, the generated shape will be sent to all the other robots through local communication so that they can merge to this shape automatically.
In the framework, the transcription factors (TFs) are used to denote the input of the GRN framework: TF1 measures the minimal distance from the current robot to the nearest obstacle and TF2 is used to maintain the number of robots.
Two genes, G1 and G2, can be thought as the processors of the robots: they are responsible to process the inputs of the GRN-framework and send signals to trigger the outputs.
Three proteins represent the output (actions) of the framework: P1 grow into an area; P2 skip an area and try to grow into another area; and P3 stop growing.
Five basic network motifs, that represent the regulations (building blocks) to constructing the GRN-framework, are proposed: \textit{positive}, \textit{negative}, \textit{OR}, \textit{AND} and \textit{XOR} \footnote{The precise mathematical formulation of these types of regulations can be found in the paper \cite{meng2012evolving}.}.
Then, using the predefined network motifs as building blocks, an evolutionary algorithm is applied to evolve structure and parameters of the GRN-framework.
In this manner each link in the GRN framework can be modeled by one of the basic network motifs.
By means of the evolutionary algorithm we need to optimize the parameters of the general GRN-framework in order to instantiate a GRN framework able to generate suitable shapes for swarm robots to adapt to unknown environments (therefore the fitness function depends on the distance from nearest obstacle and from the number of the robots within the shape).
The evolved GRN framework, presented in this paper, is able to generate the final target shape, starting from a single robot, for different environments.
Moreover, a case study is conducted in which the robots have to traverse a complex unknown environment with different constraints along the path.
The swarm of robots was able to adapt its shapes, using the GRN-based framework developed, during the traverse of a complex environment.

In the paper ``Evolving Hierarchical Gene Regulatory Networks for Morphogenetic Pattern Formation of Swarm Robots'' \cite{oh2014evolving} it is presented an approach to pattern formation for swarm robots, inspired by biological morphogenesis, that uses a hierarchical gene regulatory network (EH-GRN) evolved using network motifs.

An interesting European project relating to swarm of GRN-controlled agents whose goal is collectively organise themselves into complex spatial arrangements is Swarm-Organ (visit \url{http://www.swarm-organ.eu/}).

\subsection{GRN-based Models for Co-Evolving Body and Brain}\label{coevolve_complete}
\begin{figure}[h!]
	\centering
	\includegraphics[scale=0.5]{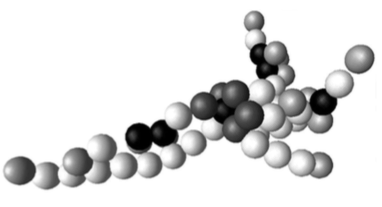}
	\caption{An example of evolved agent for block pushing. Image taken from \cite{bongard2003evolving}.}\label{fig:coevolve}
\end{figure}

In ``Evolving Complete Agents using Artificial Ontogeny'' \cite{bongard2003evolving} it is presented an artificial evolutionary system, \textit{Artificial Ontogeny} (AO), that combines an ontogenetic development with a genetic algorithm in order to evolve complete agents, that is both the morphologies and controllers of robots.
Each genome, evolved using a genetic algorithm, is treated as genetic regulatory networks, in which genes produce gene products that either have a direct phenotypic effect or regulate the expression of other genes.
In this model there is a translation from a genome (genotype) into a three-dimensional agent (phenotype), later evaluated in a physically-realist virtual environment, that takes place via ontogenetic processes: the differential gene expression and the diffusion of gene products transforms a single structural unit into an articulated three-dimensional multi-unit agent composed of several structural units that can contain sensors, actuators and a neural network structure.
Therefore, each agent begins its ontogenetic development as a single structural unit; structural units (spheres) which are the basic building blocks from which the agent's morphology is constructed.
Depending on the concentrations of gene products within a unit, the unit may grow in size and even split into two units.
Each structural unit contains at most six joints (to which other units can attach to them), a copy of the genome and six diffusion sites.
Each diffusion site contains zero or more diffusing gene products and zero or more sensor, motor and internal neurons.
Three types of sensor can be embedded, by the artificial evolution, within a structural unit: touch sensors, proprioceptive sensors and light sensors.
The neurons at a diffusion site may be connected to other neurons within the same unit or in other units.
After a unit splits from its parent unit, the two units are attached with a rigid connector.
In addition to the morphology of the agent, neural structure may grow within the developing agent.
Each genome of population is represented by 100 floating-point values (between 0.00 and 1.00) and is scanned by a parser in order to find the promotors; promotor sites indicate the starting position of a gene along the genome.
During the growth phase, the genes may emit gene products: the gene products are treated as chemicals which spread to neighbouring diffusion sites, and to a lesser degree, into neighbouring structural units.
There are 24 different types of gene products: 2 affect the growth of the unit in which they diffuse, 17 affect the growth of the agent's neural network and 5 have no phenotypic effect, but rather may only affect the expression of other genes (enhance or repress).
In the AO a cellular encoding has been incorporated to achieve the correlated growth of morphology and neural structure.
Cellular encoding is a method for evolving both the architecture and synaptic weights of a neural network, by starting with a simple neural network (embedded in each new structural unit) and iteratively applying a set of graph rewrite rules to transform it into a more complex network.
If the concentration of one of the 17 gene products, responsible for neural development, at a diffusion site exceeds a concentration of 0.8, and there is a neural structure at that site, the corresponding rewrite rule is applied to the neural structure.
This neural development scheme is able to evolve dynamic, recurrent neural network that propagate neural signals from sensor neurons to motor neurons.
In order to evolve the complete agents a genetic algorithm (with mutation and crossover) is applied with 200 generations and a population size of 200.
Each genome is evaluated (according to a task-specific fitness function) as follows: the genome is copied into a single structural unit and placed in a virtual, three-dimensional environment; morphological and neural development is allowed to proceed for 300 time steps; after this the neural network is activated and the agent is allowed to operate in its noisy environment for 1000 time steps.
The agent is the regrown and re-evaluated nine more times, and the agent's fitness values are averaged.
Using the AO system, agents able to perform directed locomotion and block pushing (see Figure \ref{fig:coevolve} for an example) in a noisy environment were evolved.

Another example of a GRN-based evolution of complete autonomous agents can be found in \cite{dellaert1996developmental}; but in this last model the nervous system develops after the development of the agent morphology.
	
\section{Morphogenetic Robotics}
\begin{figure}[h!]
	\centering
	\includegraphics[scale=0.4]{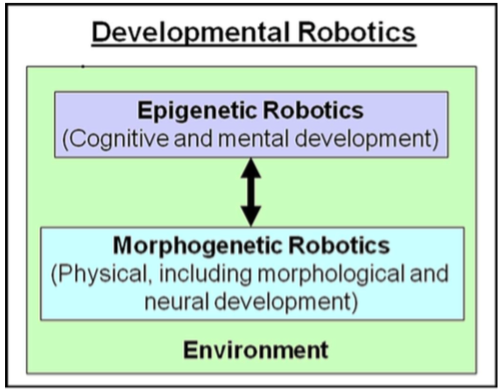}
	\caption{Morphogenetic and Epigenetic Robotics are closely coupled not only directly in that the body plan and nervous system are the basis of cognitive development, but also indirectly through the environment. Image taken from \cite{jin2011morphogenetic}.}\label{fig:morp}
\end{figure}

The term \textbf{morphogenetic robotics} has been first introduced in the paper ``Morphogenetic Robotics: An Emerging New Field in Developmental Robotics'' \cite{jin2011morphogenetic}.
Morphogenetic robotics is an emerging new field in developmental robotics that consist of a class of methodologies in robotics for designing self-organising, self-reconfigurable and self-repairable single or multi robot systems, \textit{using genetic and cellular mechanisms governing biological morphogenesis} \cite{jin2011morphogenetic}.
Biological morphogenesis is the biological process in which cells divide, grow and differentiate, and finally resulting in the mature morphology of a biological organism. Morphogenesis is under the governance of a developmental gene regulatory network and the influence of the environment \cite{gilbert2000developmental}.
They categorize these methodologies into three areas:
\begin{itemize}
	\item \textit{morphogenetic swarm robotic systems}: deal with the self-organization of swarm robots using genetic and cellular mechanisms underlying the biological early morphogenesis;
	\item \textit{morphogenetic modular robots}: modular robots adapt their configurations autonomously based on the current environmental conditions using morphogenetic principles;
	\item \textit{morphogenetic body and brain design for robots}: include the developmental approaches to the design of the body or body parts, including sensors and actuators and/or design of the neural network-based controller of robots.
	The neural structure is the product of neural morphogenesis (\textit{neurogenesis}).
\end{itemize}
The authors claim that the \textit{developmental robotics} should include both \textit{morphogenetic robotics} and \textit{epigenetic robotics}
\footnote{For a comprehensive survey of Epigenetic Robotics, and more in general, of Developmental Robotics see \cite{lungarella2003developmental}.}: the first is mainly concerned with the physical development of the body and neural control, whereas the second focuses on the cognitive and mental development.
The body morphology, as well as the neural structure of the robots is a result of morphogenetic development, on which mental development is based through interaction with the environment \cite{jin2011morphogenetic}.
In Figure \ref{fig:morp} we can see the relationship between morphogenetic robotics, epigenetic robotics and developmental robotics.

The authors of this paper introduce these three categories; below briefly summarized.

\paragraph{Morphogenetic swarm robotics}
A swarm robotic system is a multi-robot system consisting of a large number of homogeneous simple robots.
In order to apply genetic and cellular mechanisms in biological morphogenesis to self-organized control of swarm robots, it is necessary establish a metaphor between a cell and a robot.
The movement of each robot can be modelled by the regulatory dynamics of a cell.
In particular Guo Meng and Jin (in some works \cite{guo2009cellular}, \cite{meng2013morphogenetic}) have described the movement dynamics of each robot by means of a GRN model, where the concentration of two proteins represents the position of a robot and the concentration of another protein represents its velocity.
In this gene regulatory model, the target shape information is provided in terms of morphogen gradients.
This morphogenetic approach to swarm robotic systems has the advantage that the target shape can be embedded in the robot dynamics in the form of morphogen gradients, in this way the GRN model can generate implicit local interactions rules automatically to generate the global behaviour.
In addition, this model is robust to perturbations in the system and in the environment.

Others examples of morphogenesis-inspired models for swarm robotics is presented in \ref{network_motifs}.

\paragraph{Morphogenetic modular robots}
Self-reconfigurable modular robots consist of a number of modules.
They are able to adapt their shape by rearranging their modules to changing environments.
Each module has its ``body'' and its controller and each can be seen as a cell.
In fact, there are similarities in control, communication and physical interactions between cells in multicellular organisms and modules in modular robots.
The control, in both cases, is decentralized and the global behaviour emerges through local interactions of the units.
Therefore, it is a natural idea to develop control algorithms for self-reconfigurable modular robots using biological morphogenetic mechanisms \cite{jin2011morphogenetic}.
The authors present an example, taken from \cite{meng2010morphogenetic}, of morphogenetic approach to designing control algorithms for reconfigurable modular robots.
Similar to morphogenetic swarm robotic systems, each unit of the modular robot contains a chromosome consisting of several genes that can produce different proteins; the proteins can diffuse into neighboring modules.
The target configuration of the modular robot is also defined by morphogen gradients.
Morphogen gradient that each module is able to modify in order to attract or repel neighboring modules and so adapt the global configuration to the environment or task.
The attraction and repellent behaviour of the modules are regulated by a GRN-based controller.
Particularly, it is used a hierarchical approach to self-reconfiguration of modular robots: one layer defines the desired configuration of the modular robots while the other layer organizes the modules autonomously to achieve it.
This hierarchical structure is similar to those of the biological gene regulatory networks \cite{jin2011morphogenetic}.
This hierarchical controller, inspired by the embryonic development of multi-cellular organism, is resulted efficient and robust in reconfiguring modular robots to adapt to the changing environment.

\begin{SCfigure}[][h!]
	\centering
	\includegraphics[width=0.35\textwidth]{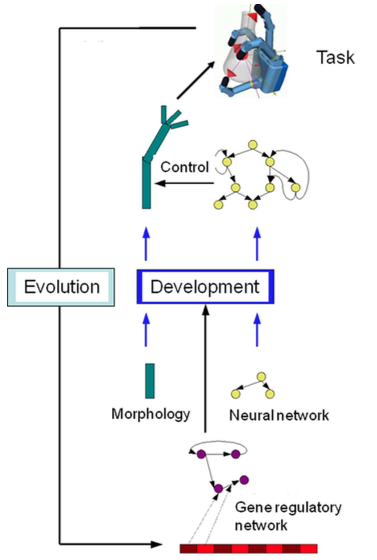}
	\caption{Conceptual diagram for coevolving the development of arm and control. Image taken from \cite{jin2011morphogenetic}.}\label{fig:hand}
\end{SCfigure}

\paragraph{Morphogenetic body and brain design for robots}
This category, according the authors, comprises models for neural \footnote{The analogy between neural development and biological neurogenesis is here reported but there is no a truly morphogenesis process because the evolution of the formal model which controls the robot is simulated.} and morphological development in designing intelligent robots.
It is also important reproduce the natural co-evolution of development of body and brain in which the cognitive and mental development is influenced by the morphological development (and by the environment) and vice versa.
In the paper \cite{jin2011morphogenetic} it is cited an example of co-evolution in development of robot hand morphology and controller, see Figure \ref{fig:hand}.
In this way the shape, the number of fingers and finger segments can be evolved together with their controller in a task-dependent way: different hand morphologies will emerge by evolving the system for different behaviours \cite{jin2011morphogenetic}.

Another example of morphogenesis-inspired co-evolution of body and brain is presented in \ref{coevolve_complete}.

An example of a morphogenesis-inspired body development is presented in \ref{evolving_eggenberger}.

\section{Conclusion}\label{conclusion}
This work represents a survey of the most relevant examples in the literature concerning the application of genetic regulatory network models in robotics.
Examples of GRN-based models are presented for designing robot control, for evolving robot morphology, for pattern generation in swarm robotics and for co-evolving body and brain.
Moreover, an introduction to another kind of classification, inspired by the biological morphogenesis, is given with the Morphogenetic Robotics.

Some of the introduced models are biologically plausible, in other words the ideas introduced in the developed artificial system can be equated with the relative biological concepts.
Others instead exploit the similarities with the cell's mechanisms but they prefer to be more abstract and therefore more computationally tractable. 
The majority of the examples found in the literature concerns the design of robot controller, and this is reflected in the paper.
This is so far due, in part, to the lack of technological support (modular robotics is an example) and to the needed computational resources (see co-evolution of the body and brain).

Nevertheless, this approach has proven to be an efficient tool for the evolution of robot capable of interesting not trivial behaviours and it represents a field of study with not yet fully explored potential.

\end{document}